# GENERAL RELATIVITY AND QUANTUM MECHANICS IN FIVE DIMENSIONS

Paul S. Wesson[1,2]

1. Department of Physics and Astronomy, University of Waterloo, Waterloo, Ontario N2L 3G1, Canada

2. Herzberg Institute of Astrophysics, National Research Council, Victoria, B.C. V8X 4M6, Canada

Abstract: In 5D, I take the metric in canonical form and define causality by null-paths. Then spacetime is modulated by a factor equivalent to the wave function, and the 5D geodesic equation gives the 4D Klein-Gordon equation. These results effectively show how general relativity and quantum mechanics may be unified in 5D.

Key Words: Higher dimensions, General relativity, Wave function, Causality

Correspondence: mail to (1) above; email = psw.papers@yahoo.ca

# GENERAL RELATIVITY AND QUANTUM MECHANICS
# IN FIVE DIMENSIONS

1. <u>Introduction</u>

The goal of unifying general relativity and quantum mechanics has a long history, but to date no approach has met with universal acceptance. Recently, however, it has been shown that all vacuum solutions of Einstein's equations can be embedded in five-dimensional canonical space (as opposed to 5D Minkowski space), and that null-paths in 5D correspond to the timelike paths of massive particles in 4D. Here I will show that these conditions lead to the modulation of spacetime by a conformal factor equivalent to the wave function, and the reduction of the 5D geodesic equation to the 4D Klein-Gordon equation of wave mechanics. Together, these technical results effectively demonstrate how 5D relativity can lead to the unification of general relativity and quantum mechanics.

A common view about unification is that we should first develop a quantum theory of gravity and then marry this to the quantum field theory of particles. The first part of this scheme has, of course, proven elusive. It therefore makes sense to consider the alternative approach, where general relativity is extended to $N(>4)$ dimensions, and the extra effects are identified with the interactions of particles. This can be done either by describing the universe by a simple extension of Einstein's theory, or by regarding our universe as one of an ensemble. In this regard should be mentioned the original extension of spacetime from 4D to 5D by Kaluza and Klein [1], which was applied to particle physics and cosmology by Dirac and Robertson [2], plus the various models of the multiverse as recently reviewed by Carr [3]. There are problems with all of these approaches, which on inspection can be traced to difficulties with choosing an appropriate base-space and a compatible definition of causality. These twin difficulties have recently been resolved, at least in principle, for the 5D case. Thus, the default metric is not the Minkowski one ($M_5$), but rather the warp metric or the canonical metric ($C_5$), which can account for particle masses and the origin of matter [4]. Below, I will use the canonical form, because it follows from Campbell's theorem that all solutions of Einstein's 4D field equations with vacuum but a finite cosmological constant can be embedded in such a 5D space. Campbell's theorem is essential to the model given below, because it enables a specific metric to be chosen for unification from among an infinitude of possibilities [5]. The other essential component of the model is the use of a 5D *null*-path to correspond to the 4D timelike path of a massive particle [6]. This means that a particle is photon-like in 5D, its mass arising dynamically and being basically 4D in nature. The 5D postulates of the canonical metric and the null-path fit well together, and then consequences can be evaluated using other technical results on the application of conformally-related spaces to mechanics [7-10]. In this way, it is now feasible to construct a model where (vacuum) general relativity and (old) quantum mechanics are unified in five dimensions.

The present account is aimed at showing how 5D classical theory can yield the wave function and the equation of motion for 4D quantum theory. I intend in the future to give a more detailed model for a particle in 5D and investigate wave-particle duality, quantization and uncertainty in 4D. The main results for the present work are derived in Sections 2 and 3. The major features of the theory and its implications for tests are summarized in Section 4.

2. 5D and the 4D Wave Functions

The model uses a 5D line element given in terms of the metric tensor by $dS^2 = g_{AB}dx^A dx^B$, where $A,B = 0,1,2,3,4$ for time, space and the extra dimension. The 4D line element $ds^2 = g_{\alpha\beta}dx^\alpha dx^\beta$ is embedded, with $\alpha,\beta = 0,1,2,3$. Since the 5D null-path given by $dS^2 = 0$ encompasses the finite 4D path $ds^2 \geq 0$ [6], the embedding necessarily involves a relation of the form $x^4 = x^4(x^\alpha)$. This defines both a hypersurface and an orbit. Since causality is to be defined via the 5D null-path, and since observations are normally made in terms of the 4D proper time, the orbit will be of the form $x^4 = x^4(s)$. The role played by the extra coordinate is seen to be different from that in other applications, being a function rather than just a distance measure. Therefore, to avoid confusion, it is advisable to revert to an old (but fortuitous) usage and label $x^4 = \psi$ [9]. The function $\psi = \psi(x^\alpha)$ can be determined once a 5D metric is chosen, which in turn depends on what 4D physics it is desired to recover on the 4D hypersurface of spacetime.

Gravitational physics near a large mass is usually discussed in terms of the Schwarzschild solution. This is a member of the larger class, comprising all vacuum solutions (possibly with a cosmological constant) of Einstein's equations [10]. For brevity, let these 4D vacuum metrics be denoted $V_4$. Also known are many solutions of the corresponding 5D field equations, when the metric is written in the so-called canonical form [8]. In this, the 5 degrees of coordinate freedom are employed to set $g_{4\alpha} = 0$ and $|g_{44}| = 1$, and the 4D part of the metric is factorized by a quadratic term in $x^4$. Canonical space $C_5$ should not be confused with 5D Minkowski space $M_5$, the difference being analogous to that between 4D metrics expressed in rectangular and spherical coordinates. The class of metrics $C_5$ is algebraically general; but when $g_{\alpha\beta} = g_{\alpha\beta}(x^\gamma \text{ only})$, the metric takes the pure-canonical form $C_5^*$ which is special (though still broad). Field equations are not central to the model to be developed, but metrics are of critical importance because they determine the dynamics (see below). It is fortunate that following Campbell there exists the *Theorem: Any member of the class $V_4$ can be locally embedded in $C_5^*$* [5,8]. This provides a firm foundation for the dynamics in the macroscopic sector of the model, as embodied in the geodesic equation.

The situation in the microscopic sector is less clear, particularly in regard to the behaviour of the wave function associated with a test particle when an observation is made (see below). But there is consensus that when the wave is propagating it is described by the Klein-Gordon equation. In textbooks, the latter is commonly derived by applying deBroglie operators to the energy-momentum-mass relation $E^2 - p^2 = m^2$. [See e.g. ref. 8 p. 25; units are chosen such that the speed of light, Newton's gravitational constant and Planck's quantum of action are all unity.] The last relation is itself based on the normalization condition for the 4-velocities $u^\alpha \equiv dx^\alpha/ds$ in the metric, namely $u^\alpha u_\alpha = 1$ or 0, depending on whether the test particle is massive or massless. The connection is via definitions for the energy $E(\sim u^0)$ and the 3-momenta $p^{123}(\sim u^{123})$ in terms of the 4-velocities, where the whole relation is multiplied throughout by $m^2$. This last procedure means that the relation $E^2 - p^2 = m^2$ actually holds irrespective of whether the mass is constant (as in Einstein theory) or variable (as in certain versions of Kaluza-Klein theory). This fact is, however, frequently obscured in particle physics, where $E$ and $p$ are treated as prime data. Notwithstanding this, the phase factor in the wave function for a free particle is the standard $(Et - px)$, using a symbolic form for the 3D properties. This factor is physically dimensionless, of course, when divided by Planck's constant. The phase factor is often written more simply as $s/\lambda$ in terms of the 4D proper time and the Compton wavelength of the particle. (The correspondence in the constant-velocity case is just a consequence of the relationship between the proper time $s$ and the coordinate time $t$ as given by the Lorentz-Fitzgerald time-dilation formula.) The reason for recalling these things is that, for the wave sector of the model, it is necessary that the Klein-Gordon equation be recovered with the appropriate phase factor.

The comments of the two preceding paragraphs actually constrain the model strongly when combined with the demand that the 5D interval be null. In fact, the essential physics follows automatically from this assumption plus the requirement that the 5D metric have the pure-canonical form. Thus:

$$dS^2 = 0 = (\psi/L)^2 g_{\alpha\beta}(x^\gamma) dx^\alpha dx^\beta + d\psi^2 \qquad (1.1)$$

$$= (\psi/L)^2 ds^2 + d\psi^2 \qquad . \qquad (1.2)$$

Here the $C_5^*$ quadratic factor has been written as a ratio of two lengths, because the constant length $L$ sets a scale for the geometry of the 4-space and will turn out to have a distinct meaning. The quadratic factor gives the 5D metric the form of the 4D synchronous metric, and means that the 5D space contains a kind of spherical 4D subspace. The extra dimension in (1) is taken to be timelike, since both signs are allowed in modern Kaluza-Klein theory. This choice means that the cosmological constant associated with the 4D part of (1) is $\Lambda < 0$ and that spacetime is closed. These comments follow from the vacuum Einstein equations $G_{\alpha\beta} = \Lambda g_{\alpha\beta}$, where the 4D scalar curvature is $R = -G = -4\Lambda$ [4, 7]. There is no problem with closed timelike paths in (1), because the extra coordinate does not have the physical nature of a time (see ref. 8; an alternate route is to take a spacelike extra dimension and apply a Wick rotation later). From (1) there comes

$$\frac{d\psi}{ds} = \pm \frac{i\psi}{L} \qquad (2.1)$$

$$\psi = \psi_* e^{\pm is/L} \qquad (2.2)$$

The sign choice here reflects the reversibility of the 'velocity' in the extra dimension, and $\psi_*$ is a constant amplitude. Incidentally, an attempt to add a constant shift to (2.2) via $\psi \to (\psi - \psi_0)$ results in a divergence in the 4-geometry similar to that of the singular hypersurface of membrane theory [4], so this possibility is postponed to future work. Obviously, the main import of (2.2) is that it describes an oscillation around $\psi = 0$ with wavelength $L$. This should be identified with the Compton wavelength ($1/m$) of the associated particle if the wave has energy given by Planck's law. Then (2.2) is essentially the standard wave function.

This conclusion is supported by a more detailed investigation. The 5D null-path assumption (1) is a scalar statement. It is true that it yields the path or orbit (2.2) in a direct fashion, analogous to how the path of a light ray is obtained from the metric in 4D cosmology. However, as in that situation, more information is obtainable by considering the variation of the proper time around its mean value, to obtain the extremum. In the present situation, this procedure implies

$$\delta\left[\int dS\right] = 0 \quad , \tag{3}$$

which can be taken about $dS = 0$. The result is the 5D geodesic equation, which can be expressed in terms of the 4D proper time $s(\neq 0)$. The 5D equations of motion split naturally into a 4-component set for spacetime, plus a component for the extra dimension:

$$du^\mu/ds + \Gamma^\mu_{\alpha\beta} u^\alpha u^\beta = 0 \tag{4.1}$$

$$\frac{d^2\psi}{ds^2} - \frac{2}{\psi}\left(\frac{d\psi}{ds}\right)^2 - \frac{\psi}{L^2} = 0 \quad . \tag{4.2}$$

The first set here shows that the 4D part of the 5D motion is *identical* to the geodesic of standard general relativity, where the Christoffel symbol accounts for the curvature. This remarkable result can be traced to the fact that $\partial g_{\alpha\beta}/\partial\psi = 0$ in the metric (1.1), so the motion in spacetime is decoupled from that in the extra dimension. The equation of motion (4.2) for the extra dimension looks at first glance to describe simple-harmonic motion with a velocity-dependent friction term. However, it is solved by the simple wave (2.2) found from the metric. To further understand this, it is instructive to carry out the duality transformation $\psi \to L^2/\psi$ in the metric (1.2) and the equation of motion (4.2). The former loses the $C_5^*$ form, but the latter is still a valid form because the relation concerned is tensor-derived and therefore covariant (see ref. 4 concerning this transformation). The new forms are:

$$dS^2 = 0 = (L/\psi)^2 ds^2 + (L/\psi)^4 d\psi^2 \tag{5.1}$$

$$\frac{d^2\psi}{ds^2} + \frac{\psi}{L^2} = 0 \quad . \tag{5.2}$$

The latter equation, unlike (4.2), manifestly describes simple-harmonic motion. If the problem were a mechanical one, the 'spring constant' would be $1/L^2$. Now recalling that $L$ is the Compton wavelength, the restoring constant is just $m^2$, where $m$ is the rest mass of the particle associated with the wave. The wave must be supported by the vacuum. As for (4.2), the relation (5.2) is solved by (2.2) or $\psi = \psi_* e^{\pm ims}$.

We see that the 5D metric and its associated equations of motion both imply that the conformal factor typical of 5D canonical space has the physical meaning of the 4D wave function.

3. <u>5D and the 4D Klein-Gordon Equation</u>

When operators are applied to the 4D metric, the result is the relativistic wave equation called after Klein and Gordon. The non-relativistic version of this is the Schrodinger equation. The Klein-Gordon equation is a scalar relation. By comparison, in the new approach being investigated here, the scalar relation (3) is in effect a constraint on the tensor relations (4). This implies that the spacetime components (4.1) are in some sense equivalent to the extra component (4.2); and that since (4.1) are the same equations of motion as in standard theory, (4.2) must in some sense be equivalent to the Klein-Gordon equation.

This may indeed be shown with some algebra. It is convenient in this context to use a comma to denote the ordinary partial derivative and a semicolon to denote the regu-lar (4D) covariant derivative. Then the geodesic equations (4.1) for spacetime became $u^\beta u^\alpha_{;\beta} = 0$, with summation as elsewhere. It is also convenient to define $g^{\alpha\beta}\psi_{,\alpha;\beta} \equiv \Box^2 \psi$. Then the geodesic for the extra dimension (4.2) may be expanded using $d\psi/ds = \psi_{,\alpha}u^\alpha$ etc., and a term set to zero by using (4.1). On replacing $L$ with $1/m$ by previous considerations, (4.2) may be written

$$\Box^2\psi + m^2\psi = 0 \quad . \tag{6}$$

This is the standard Klein-Gordon equation, here derived from the equation of motion for the extra dimension of Kaluza-Klein theory assuming that the 5D path is null.

Einstein's field equations can be used carry out further investigations. This because (1) is algebraically tantamount to creating a new metric which is conformally related to the old one. Many results are known about conformally-related metrics, both in 4D [7, 10] and 5D [8, 9; for what follows, see especially ref. 7 p. 446 and ref. 8 p. 236]. Employing these results, the components of the Riemann tensor, the Ricci scalar and the components of the Einstein tensor can be evaluated in the new frame from their values in the old one. However, since a conformal transformation is in general not equivalent to a coordinate transformation, the expressions in the new frame are expected to be quite different to the ones in the old frame. This is particularly relevant to the Einstein tensor and the energy-momentum tensor, which in both frames are related by $G_{\alpha\beta} - \Lambda g_{\alpha\beta} = 8\pi T_{\alpha\beta}$. Metrics with $C_5^*$ form obey these equations with a 'source' that is a vacuum fluid with the equation of state $p = -\rho = -\Lambda/8\pi$, where $\Lambda$ is the cosmological constant. The way to determine the properties of the model in the conformal frame is to carry out the trans-formation implied by (1). An overbar will denote quantities in the conformal frame. For the Ricci tensor:

$$\bar{R}_{\alpha\beta} = R_{\alpha\beta} - \frac{2\psi_{,\alpha;\beta}}{\psi} + \frac{4\psi_{,\alpha}\psi_{,\beta}}{\psi^2} - \left(\frac{\psi^{,\gamma}_{;\gamma}}{\psi} + \frac{\psi^{,\gamma}\psi_{,\gamma}}{\psi^2}\right)g_{\alpha\beta} \quad . \tag{7}$$

In this, it may be tempting to set the term in parentheses to zero, because it is a quasi-Klein-Gordon equation if the mass is rescaled $(m^2 \rightarrow m^2/\psi^2)$ and because it is one of two solutions to the extra field equation in the 5D analog of the present 4D problem ($R_{44} = 0$; see ref. 8 p. 236). But in the present problem this would represent an extra assumption, so it is avoided. It is better to contract (7) without resort to (6), giving the Ricci scalar:

$$\bar{R} \equiv \bar{g}^{\alpha\beta}\bar{R}_{\alpha\beta} = \frac{L^2}{\psi^2}\left(R - \frac{6\psi^{,\gamma}_{;\gamma}}{\psi}\right) \quad . \tag{8}$$

If (6) is now employed, a convenient relation is obtained between the curvature scalars in the conformal and original spaces in terms of the mass:

$$\bar{R} = (L/\psi)^2 (R + 6m^2) \quad . \tag{9}$$

This relation is mathematically simple but physically important. Notably, the starting metric (1) implies a spacetime which is empty of ordinary matter but has a cosmological constant $\Lambda$, so Einstein's equations imply that this contribution is measured by $R = -4\Lambda$. Then (9) states that the scalar curvature of the conformal space measures the sum of the energies associated with the vacuum and the mass of the particle. This is eminently sensible. In other applications, it may be useful to have the components of the Einstein tensor for the conformal frame which accounts for both the wave/particle and the background spacetime. Equations (7) and (8) can be employed to form $\overline{G}_{\alpha\beta} \equiv \overline{R}_{\alpha\beta} - (\overline{R}/2)\overline{g}_{\alpha\beta}$ with the result

$$\overline{G}_{\alpha\beta} = G_{\alpha\beta} - \frac{2\psi_{,\alpha;\beta}}{\psi} + \frac{4\psi_{,\alpha}\psi_{,\beta}}{\psi^2} + \left( \frac{2\psi^{,\gamma}_{;\gamma}}{\psi} - \frac{\psi^{,\gamma}\psi_{,\gamma}}{\psi^2} \right) g_{\alpha\beta} \quad . \tag{10}$$

This is the general result for manifolds related by the starting assumption (1). It should be noted that (10) gives $\overline{G} \equiv \overline{G}^{\alpha}_{\alpha} = (L/\psi)^2 (G - 6m^2)$, in agreement with (9). It should also be noted that the analysis of this paragraph mainly involves two conformally-related *four*-dimensional spaces, and does not depend directly on the fifth dimension. But while equations (7)-(10) hold in 4D, they follow logically from the assumption (1) of a null interval in 5D.

In other words, the 4D Klein-Gordon equation and the other relations derived above rest on the 5D canonical metric and its associated null-path.

4. <u>Conclusion and Discussion</u>

The existence of even one extra dimension over the four familiar ones of spacetime is yet unproven. In order to be considered "real", there has to be some prospect of communicating in N(>4)D. This needs a workable definition of causality, which is compatible with the 4D one where photons move on paths where the element of proper time obeys $ds^2 = 0$. The logical extension is to higher-dimensional paths where *all* particles (including massive ones) move on null-paths with $dS^2 = 0$. This postulate, combined with the canonical metric, leads readily to the 4D wave function and the 4D Klein-Gordon equation.

The relative ease of the derivations in Sections 2 and 3 deserves comment. It is clear that the 5D canonical metric $C_5$ is much easier to deal with than the 5D Minkowski metric $M_5$ or its generalized forms, a difference analogous to how in 4D many natural systems are easier to describe with metrics whose 3D sections have spherical as opposed to rectangular symmetry. The condition that the 5D path be null makes the approach even easier. Indeed, the chain of logic is straightforward: The 5D null-path (1) allows massive particles to have 4D timelike paths, where the 5D space can by Campbell's theorem be taken to be the pure canonical one which embeds the 4D Schwarzschild and other vacuum solutions of general relativity, such that the prefactor on the 4D subspace has wavelike properties, with dynamical relations (2)-(5) which lead to the Klein-Gordon equation (6), while more insight is forthcoming by applying Einstein's equations to the kernel space and the related conformal space to obtain equations (7)-(9), which imply the relationship (10) between the Einstein tensors, thus completing the inventory of 4D information as derived from 5D.

The model given here can clearly be extended to any number $N$ of dimensions. There are many versions of the universe with N(>4)D, and of multiple universes [3]. A motivation often quoted in support of these models is that they can resolve, at least in principle, the information-loss problem which occurs in general relativity, when a complicated object falls into a black hole, for example. The resolution is assumed to be that information 'lost' in 4D is encoded in the fields of the fifth or higher dimensions. This sounds plausible, and may occur in the model presented above, where causality is defined via $dS^2 = 0$ and cannot be violated in a 5D sense. However, it must be admitted that information loss, via the breakdown of unitarity, afflicts many formulations of the physics of microscopic as opposed to macroscopic systems. Even the Klein-Gordon equation has problems in that some solutions to it cannot be consistently interpreted as relativistic wave functions, so unitarity breaks down and information is apparently lost. This problem is compounded by the different roles played by parameters and operators in quantum mechanics, and in particular by the different nature of the time and the spatial coordinates. This has led some workers to speculate that a problem-free unification of classical and quantum physics may involve quantum field theory in distinction to wave mechanics. In response to this, some proponents of N-dimensional unification have argued that a true marriage of pure quantum fields is beyond reach at present, and that a practical approach is via the intermediate step of a unification of classical gravity with wave mechanics. This is essentially what I have carried out here.

The comparative ease with which unification can be carried out in 5D, using the canonical metric and the null-path, is in my view an indication of the basic correctness of the approach adopted here. That said, I realize that the approach needs to be both refined and generalized. This is especially true in order to formulate tests and predictions of the model. There are quite a number of these. Specifically: (a) Higher-dimensional field equations were not employed above, where the emphasis was on metric-based dynamics, because of a lack of consensus. The only shared belief among workers appears to be that the $N$D field equations should involve the Ricci tensor $R_{AB}(A, B = 0, 123, 4...)$. Even with one extra dimension, and agreement about the mathematical structure of the field equations, there is controversy about their physical application. Some use the canonical metric as a basis for explaining 4D matter as a consequence of 5D geometry, while others use the warp metric as a means of splitting the manifold with a singular membrane and so explaining the interactions and masses of particles [4]. Both approaches agree with observations. This success is due ultimately to Campbell's theorem, which ensures that the 5D $R_{AB}$ contains the 4D $G_{\alpha\beta}$ [5, 8; in fact $R_{AB} = 0$ for apparent 5D vacuum implies Einstein's 4D equations *with* matter]. The inference is that for $N \geq 5$, Campbell's theorem should guide the choice of both metric and field equations. The latter can in principle be used, as noted at equation (7) above, to simplify the analysis of the spacetime embedding.

(b) Topology was not considered above, again because of a lack of consensus. Any field equations of classical type do not inform directly about topology, which must be fixed either by boundary conditions (or the lack thereof as in Einstein's closed cosmological model) or by quantum considerations (as in Klein's electron model). It is popular to take the extra dimension of modern Kaluza-Klein theory to be noncompact; but the higher dimensions needed to incorporate the symmetry groups of elementary particles may still be compact, as might in principle be the regions of spacetime associated with them. Symmetries should obviously be used judiciously in simplifying the metric of any *N*D model.

(c) Quantization and uncertainty were not explicitly mentioned above, partly because the existence of a wave implies the discretization of energy and the non-localizability of the associated particle. It is close to trivial to generalize the wave described above to include overtones measured by an integer *n*, at least in a mathematical sense. There may, how-ever, be more to this in a physical sense. For example, in a certain interpretation of the canonical metric, it is straightforward to show that the standard quantization rule $\int mds = n$ is equivalent to the trapping of the particle in a higher-dimensional space with structure [4,5]. In general, there are questions to do with wave-particle duality, quantization and uncertainty which I expect to report on in future. (d) Matter is absent in the model outlined above, in the sense that the embedded Einstein space is a vacuum one. This can be most easily remedied by changing the metric from the pure-canonical one to the regular canonical one [4, 8; this means that $g_{\alpha\beta}(x^\gamma) \to g_{\alpha\beta}(x^\gamma, x^4)$ for the 4D metric tensor]. Many 5D cosmological solutions with acceptable properties of matter are already known, but the behaviour of the quantum wave function in them is largely unknown. (e) The scalar field was suppressed in the above, by setting $|g_{44}(x^\gamma, x^4)| \to 1$. This condition should be relaxed, because there is ample evidence from cosmological and other solutions that the scalar potential is associated with a kind of mass field. (f) The electromagnetic field was removed in the above, by setting $g_{4\alpha}(x^\gamma, x^4) \to 0$. If this condition were relaxed, it would be possible to investigate possible departures from Coulomb's law and variations in the fine-structure constant, in space over particle distances and in time over cosmological periods.

The comments (a)-(f) above show that the approach adopted here is fruitful and will repay further investigation. As pointed out in the first line of this paper, the unification of general relativity and quantum mechanics (in whatever form) has been a long-standing goal in physics. On reflection, it is remarkable and fortunate that unification can be achieved at all by adding just one extra dimension.

Acknowledgements

Thanks for comments go to B. Mashhoon and other contributors to the S.T.M. website at http://www.astro.uwaterloo.ca/~wesson. This work was partly supported by N.S.E.R.C.

References

[1]  T. Kaluza, Sitz. Preuss. Akad. Wiss. 23 (1921). O. Klein, Z. Phys. 37 (1926) 895.

[2]  P.A.M. Dirac, Ann. Math. 36 (1935) 657. H.P. Robertson, T.W. Noonan, Relativity and Cosmology, Saunders, Philadelphia (1968) p. 413.

[3]  B. Carr, Universe or Multiverse? Cambridge Un. Press (2007).

[4]  P.S. Wesson, Five-Dimensional Physics, World Scientific, Singapore (2006). P.S. Wesson, Gen. Rel. Grav. 40 (2008) 1353.

[5]  S. Rippl, C. Romero, R. Tavakol, Class. Quant. Grav. 12 (1995) 2411. S.S. Seahra, P.S. Wesson, Class. Quant. Grav. 20 (2003) 1321.

[6]  S.S. Seahra, P.S. Wesson, Gen. Rel. Grav. 33 (2001) 2371. D. Youm, Mod. Phys. Lett. A 16 (2001) 2371. P.S. Wesson, Class. Quant. Grav. 19 (2002) 2825.

[7]  R.M. Wald, General Relativity, Un. Chicago Press (1984).

[8]  P.S. Wesson, Space-Time-Matter, 2nd. ed., World Scientific, Singapore (2007).

[9]  J. Ponce de Leon, Gen. Rel. Grav. 20 (1988) 539.  P.S. Wesson, Astrophys. J. 394 (1992) 19.  P.S. Wesson, Phys. Lett. B 276 (1992) 299.

[10] D. Kramer, H. Stephani, M. MacCallum, E. Herlt, Exact Solutions of Einstein's Field Equations, Cambridge Un. Press, Cambridge (1980).